\documentclass[runningheads]{llncs}
\usepackage[T1]{fontenc}
\usepackage{graphicx}
\usepackage{soul}
\usepackage{hyperref}
\usepackage{amsmath}
\usepackage{amssymb}
\usepackage{xcolor}
\usepackage[english]{babel}
\usepackage{tcolorbox}

\usepackage{adjustbox}
\usepackage{booktabs}
\usepackage{multirow}
\usepackage{colortbl}

\usepackage{enumitem}

\usepackage{appendix}
\usepackage{balance}
\usepackage{colortbl}
\usepackage{amsmath}
\usepackage{booktabs}
\usepackage{tabularx}
\usepackage{multirow}
\usepackage{caption}
\usepackage{subcaption}
\usepackage{adjustbox}
\usepackage{lmodern}
\usepackage{tikz}
\usepackage{amssymb}
\usepackage{pifont}

\usepackage[table]{xcolor}

\usepackage{orcidlink}
\setitemize{noitemsep}
\setlist{topsep=0pt, leftmargin=*}

\usepackage[font=small]{caption}

\newcommand{\uls}{\begin{itemize}[leftmargin=*]}
\newcommand{\ule}{\end{itemize}}
\newcommand{\ols}{\begin{enumerate}[leftmargin=*]}
\newcommand{\ole}{\end{enumerate}}
\newcommand{\li}{\item}

\usepackage{tikz}
\usepackage{bm}

\newcommand{\topdocs}{L_\theta(q)}
\newcommand{\fusedtopdocs}{\mathcal{L}_\Theta(q)}

\newcommand{\rrf}{RRF}
\newcommand{\comb}{CSUM}

\newcommand{\standard}{\mathbb{E}_\Theta[\tau_Q]}

\newcommand{\para}[1]{\paragraph{\textnormal{\textbf{#1}.}}}

\usepackage{marginnote}

\usepackage{orcidlink}

\begin{document}
%





\title{Beyond Correlations: A Downstream Evaluation Framework for Query Performance Prediction}

\titlerunning{Downstream-aware QPP Evaluation}



\author{
Payel Santra\inst{1}\orcidlink{0009-0005-5721-248X} \and
Partha Basuchowdhuri\inst{1}\orcidlink{0000-0003-1588-4665} \and 
Debasis Ganguly\inst{2}\orcidlink{0000-0001-7655-7591}
}

\authorrunning{Santra et al.}

\institute{
Indian Association for the Cultivation of Science, Kolkata, India
\\
\email{payel.iacs@gmail.com, partha.basuchowdhuri@iacs.res.in}
\and
University of Glasgow, Glasgow, UK
\\
\email{debasis.ganguly@glasgow.ac.uk}
}
\maketitle              
\begin{abstract}

The standard practice of query performance prediction (QPP) evaluation is to measure a set-level correlation between the estimated retrieval qualities and the true ones. However, neither this correlation-based evaluation measure quantifies QPP effectiveness at the level of individual queries, nor does this connect to a downstream application, meaning that QPP methods yielding high correlation values may not find a practical application in query-specific decisions in an IR pipeline.
In this paper, we propose a downstream-focussed evaluation framework where a distribution of QPP estimates across a list of top-documents retrieved with several rankers is used as priors for IR fusion.
While on the one hand, a distribution of these estimates closely matching that of the true retrieval qualities indicates the quality of the predictor, their usage as priors on the other hand indicates a predictor's ability to make informed choices in an IR pipeline. Our experiments firstly establish the importance of QPP estimates in weighted IR fusion, yielding substantial improvements of over 4.5\% over unweighted CombSUM and RRF fusion strategies, and secondly, reveal new insights that the downstream effectiveness of QPP does not correlate well with the standard correlation-based QPP evaluation.

\end{abstract}

\section{Introduction}

Query Performance Prediction (QPP) methods aim to estimate the retrieval effectiveness of a ranked list without access to relevance judgments \cite{arabzadeh2021bert,roitman2017robust,uef_kurland_sigir10,NQC}. Unsupervised QPP approaches mostly leverage the retrieval score distribution of the top documents obtained with the original query \cite{roitman2019normalized,roitman2017robust,NQC} or its variants \cite{datta2022relative,Oleg_2019}, along with other characteristic features, such as term informativeness \cite{zhou2007query}, co-occurrence \cite{uef_kurland_sigir10} and semantics \cite{10.1145/3539618.3591625}. However, supervised approaches rely on the query content and retrieved documents to predict the retrieval quality \cite{arabzadeh2023noisy,arabzadeh2021bert,hashemi2019performance}.


In its standard form, the objective of the QPP task is to effectively distinguish easy queries from difficult ones for a given retrieval model. Accordingly, its evaluation typically focuses on measuring the correlation between the true retrieval effectiveness (e.g., AP or nDCG) and the predicted QPP scores~\cite{Ganguly2022AnAOA}. However, existing QPP models are largely agnostic to both the underlying information retrieval (IR) model and the evaluation metric. Consequently, the observed correlations are often highly sensitive to these factors, i,e., observations are mostly not consistent across a range of different ranking models and target IR metrics~\cite{Ganguly2022AnAOA}.
Another major limitation of the correlation-based QPP evaluation paradigm is its dependence on a set of queries, in contrast to ad-hoc IR evaluation, which can be performed independently for each query --- specifically, by assessing whether the retrieved documents satisfy that query’s information need. Consequently, QPP evaluation is inherently \emph{conditioned} on the query set, where the performance of one query is interpreted \emph{relative} to that of others, unlike per-query measures such as AP or nDCG that evaluate retrieval quality in isolation.

Beyond the lack of robustness and per-query independence, the most critical shortcoming of existing QPP evaluation methodologies is their misalignment with the downstream utility of QPP estimates—the very motivation behind the task. To address this, we propose a \textbf{downstream-aware QPP evaluation framework}. Given an input query and the corresponding rank list, our QPP evaluation approach first employs a QPP model to predict the expected retrieval quality. This formulation enables per-query evaluation of a QPP model by predicting a relative preference of rankers for each query,
while also facilitating a natural downstream application—using the predicted scores as weights in an information retrieval fusion framework. If the predicted likelihoods closely approximate the true retrieval performance, incorporating them as priors in the fusion process should enhance overall retrieval effectiveness. Consequently, a QPP model can be regarded as more effective than another if its estimates, when employed as priors in an IR fusion algorithm, yields more effective retrieval quality than its competitor.
In this way, we not only contribute a downstream-oriented perspective on QPP evaluation but also demonstrate that ad hoc retrieval performance can be enhanced through QPP-based fusion.
The source code for the proposed QPP evaluation framework and the QPP-based IR fusion is made available for research purposes\footnote{\url{https://github.com/payelsantra/QPP-Fusion}}.

\section{Downstream Evaluation of QPP Models on IR Fusion} \label{sec:method}
\subsection{QPP across multiple rankers}
Standard QPP is formalized as a function of the form
$\phi(q, \topdocs) \mapsto \mathbb{R}$,
which predicts the retrieval quality for a query $q$ and its top documents $\topdocs$ obtained from a single ranker $\theta$. Instead of using these estimates to distinguish between multiple queries, we rather employ a QPP model $\phi$ to obtain predicted retrieval qualities across several IR models, i.e., for a set of available rankers $\Theta = \{\theta_1,\ldots,\theta_m\}$, a QPP model outputs a distribution of estimated retrieval qualities across the rankers. Formally,
\begin{equation}
\phi(q, \fusedtopdocs) \mapsto \mathbb{R}^m, \text{where}\, \fusedtopdocs = \cup_{\theta \in \Theta}\topdocs,
\label{eq:local}
\end{equation}

\subsection{QPP for Weighted Fusion} An IR fusion model combines the top-retrieved document lists obtained from multiple rankers into a single ranked list by aggregating the scores or ranks of each document across these lists \cite{fox1994combination,rrf}. 
Formally,
\begin{equation}
\theta_f(q, d; \Theta) = \sum_{\theta \in \Theta} P(\topdocs|d)\,\sigma(q, d;\theta),
\label{eq:weighted_eq}
\end{equation}
where $\theta_f: q \times d \times \Theta \mapsto \mathbb{R}$ denotes a generic weighted fusion scoring function,
$P(\topdocs|d)$ represents the prior likelihood of selecting a particular document list conditioned on $d$, and
$\sigma(q, d; \theta) \in \mathbb{R}$ denotes the similarity scoring function of the IR model $\theta$.

A uniform prior, $P(\topdocs|d) = 1$, corresponds to the unweighted fusion method like, CombSUM \cite{fox1994combination}, while setting
$P(\topdocs|d) = \sum_\theta \mathbb{I}(d \in \topdocs)/m$ (where $m = |\Theta|$) yields CombMNZ~\cite{fox1994combination}.
These priors, $P(\topdocs|d)$, can also be optimized using a set of training queries with known relevance assessments \cite{probfuse}.

Unlike existing fusion approaches that employ uniform or maximum-likelihood based priors $P(\topdocs|d)$ as weights for each ranked list,
we propose to use QPP estimates to determine these weights. More specifically, Equation \ref{eq:weighted_eq} becomes 
\begin{equation}
\theta_f(q, d; \Theta,\Phi) = \sum_{\theta \in \Theta} \phi(q, \topdocs)\,\sigma(q, d;\theta) \label{eq:qppfusion}
\end{equation}
where $\phi(q, \topdocs)$ denotes the QPP estimate for the ranked list $\topdocs$ so that the function $\theta_f$ of Equation \ref{eq:weighted_eq} now additionally also depends on the estimated QPP distribution: $\Phi = (\phi_1,\ldots,\phi_m)$.
%
The underlying intuition is that a more accurate distribution of QPP estimates across the ranked lists---one that more closely reflects their true retrieval quality---should lead to a more effective fusion outcome. More precisely, a QPP model $\phi$ is considered to be more effective than another $\phi'$ if for a query $q$
the QPP weights obtained with $\Phi$ yield a better retrieval results, as measured with an IR metric $\mathcal{P}$ (e.g., AP, nDCG etc.), than those with $\Phi'$ for $q$.
This is clearly different from the standard downstream-agnostic based QPP evaluation mechanism, which considers a QPP model $\phi$ to be more effective than $\phi'$ if the estimates of $\phi$ over a set of queries correlate better with a target IR metric than that of $\phi'$s.

\section{Experimental Setup} \label{ss:setup}

\para{Research Questions}
We conduct our experiments on the TREC DL 2019~\cite{DBLP:journals/corr/abs-2003-07820} and 2020 \cite{DBLP:journals/corr/abs-2102-07662} datasets to explore the following RQs.
\uls 
\li[] \textbf{RQ1}: Does the usage of QPP estimates as prior weights in IR fusion improve the retrieval effectiveness of over non-weighted fusion methods? 
\li[] \textbf{RQ2}: How correlated is the downstream-aware QPP evaluation with the standard QPP evaluation? 
\li[] \textbf{RQ3}: How strongly does per-query QPP effectiveness correlate with downstream fusion gains?
\ule 

\para{Evaluation Measures}
As we employ several rankers in our fusion-based QPP evaluation setup, we report the average of the individual correlations across rankers as the standard QPP evaluation measure in our experiments. We call this correlation $\mathbb{E}_\Theta[\tau_Q]$, i.e., \textbf{correlation over queries averaged across rankers}.

%

Moreover, since in the fusion-based IR pipeline, a QPP model is used to output a distribution of performance prediction over a set of rankers (Equation \ref{eq:local}), it is possible to compute a correlation between the predicted and the true ranker preferences on a per-query basis as an adaptation of the standard QPP correlation measure suitable for our task. We call this measure $\mathbb{E}_Q[\tau_\Theta]$ to indicate \textbf{correlation over rankers averaged across queries}.




\para{IR Models Investigated} \label{ss:rankers}
Since the proposed QPP evaluation framework predicts the retrieval quality distribution across several rankers (Equations \ref{eq:local} and \ref{eq:weighted_eq}), 
we employ the following $m=8$ ranking models of different characteristics for our experiments:
1) \textbf{BM25} \cite{robertson2004understanding}, a classic term weighting model based on term overlaps, term informativeness and document length normalisation, and 2) \textbf{RM3} \cite{RelevanceFeedback} which is a pseudo-relevance feedback (PRF) method based on local term co-occurrences as representatives of \textbf{sparse} models; 3) \textbf{Splade} \cite{formal2021splade}, a representative from the \textbf{learned sparse} class that relies on contextual term expansion under sparsity regularization;
4) \textbf{E5} \cite{wang2022text}, 5) \textbf{ColBERT} \cite{colbert_sigir20}, a \textbf{bi-encoder} and a \textbf{late interaction} model, respectively, both involving semantic representation encoding as dense vectors, 6) \textbf{ColBERT-PRF} and 7) \textbf{ColBERT-PRF-reranker} \cite{wang2023colbert}, PRF approaches to enrich the ColBERT vectors for query terms --- the latter only involving a reranking step, and 8) \textbf{MonoT5} \cite{monot5}, a cross-encoder model involving a \textbf{retrieve-and-rerank} pipeline on BM25 results.
%
%
%
\para{QPP Models Investigated} \label{ss:qpp-methods}
Among \textbf{retrieval score-based methods}, we employ 1) WIG \cite{zhou2007query}, 2) \textbf{NQC} \cite{NQC}, 3) \textbf{SCNQC} \cite{roitman2019normalized}, 4) \textbf{SMV}~\cite{tao2014query}, 5) $\mathbf{\sigma_{\text{max}}}$~\cite{perez2010standard}, and 6) $n(\sigma_{x\%})$~\cite{cummins2011improved}, which, generally speaking, estimate IR effectiveness as the variance of the retrieval scores.
Among \textbf{robustness-based models}, we examine 7) \textbf{UEF} \cite{uef_kurland_sigir10} and 8) \textbf{RSD} \cite{roitman2017robust}, which estimate IR effectiveness through QPP stability of subsamples of top-documents. We also employ 9) \textbf{QV-NQC}, a query variant based approach proposed in \cite{datta2022relative,Oleg_2019}. Among neural representation based unsupervised approaches, we employ 10) \textbf{DM} \cite{10.1145/3539618.3591625} and 11) \textbf{QPP-PRP} \cite{singh2023unsupervised}, which estimate IR performance from document embedding dispersion and pairwise ranking consistency, respectively. Among supervised approaches, we experiment with feature-based
12) \textbf{NQA-QPP} \cite{hashemi2019performance} and cross-encoder based 13) \textbf{BERTQPP} \cite{arabzadeh2021bert} which learn to predict retrieval quality as a function of the content of queries and their top-retrieved documents.

\para{IR Fusion Methods Investigated} 

As baselines, we report results obtained using two standard unweighted fusion approaches, implemented as instances of Equation~\ref{eq:weighted_eq} with $P(\topdocs|d) = 1$: CombSUM (\textbf{CSUM}) \cite{fox1994combination} and reciprocal rank fusion (\textbf{RRF}) \cite{rrf}.
In the case of RRF, the reciprocal of the rank at which a document $d$ is retrieved for a query $q$ by a retrieval model $\theta$ is substituted for the score $\sigma(q, d; \theta)$ in Equation~\ref{eq:weighted_eq}. 

It is worth mentioning that we also experimented with CombMNZ \cite{fox1994combination}, where the priors are obtained with likelihood estimates. Although we observed that the results of weighted CombMNZ~(over 8 ranking models) outperformed unweighted CombMNZ results, and also QPP priors yielded better results for \comb~and \rrf. As a result we exclude the QPP-guided CombMNZ variant from our reported results.  

Additionally, to seek a stronger baseline, we also tuned the priors of each ranker by a two-fold cross validation setup over DL'19 and DL'20 topics similar to \cite{probfuse}. However, as this yielded worse results than \comb~and \rrf~(likely, due to lack of an adequate number of queries in the training fold), we do not include this weaker baseline. 

\begin{table}[t]
\centering
\small
\begin{adjustbox}{width=.99\textwidth}
\begin{tabular}{@{}l@{~~}r@{~~}r@{~~}r@{~~}r@{~~}r@{~~}r@{~~}r@{~~}r@{~~}c@{~~}r@{~~}r@{~~}r@{}}
\toprule
 & \multicolumn{6}{c}{TREC DL'19} & \multicolumn{6}{c}{TREC DL'20} \\
\cmidrule(r){2-7} \cmidrule(r){8-13}
& \multicolumn{2}{c}{AP@100} & & \multicolumn{2}{c}{nDCG@10} & & \multicolumn{2}{c}{AP@100} & & \multicolumn{2}{c}{nDCG@10} & \\
\cmidrule(r){2-4} \cmidrule(r){5-7} \cmidrule(r){8-10} \cmidrule(r){11-13}
Unweighted & \rrf & \comb & & \rrf & \comb & & \rrf & \comb &  & \rrf & \comb & \\
\cmidrule(r){2-3} \cmidrule(r){5-6}
\cmidrule(r){8-9} \cmidrule(r){11-12}
(Baseline) & .488 & \textbf{.500}& $\standard$ & .737 & \textbf{.758}& $\standard$ &.497 &\textbf{.520}&$\standard$ &.715& \textbf{.737}& $\standard$\\

\midrule
$\sigma_{\text{max}}$ & \cellcolor{green!10}.491 &\cellcolor{green!10}.509 &.299 & \cellcolor{green!10}.744 & \cellcolor{red!10}.757 & .217
& \cellcolor{green!10}.515 & \cellcolor{green!10}.520  & .250&\cellcolor{red!10}.711 & \cellcolor{red!10}.730 & .163\\
$n(\sigma_{x\%})$ & \cellcolor{green!10}.495 & \cellcolor{green!10}.505 & .302&\cellcolor{green!10} .742 & \cellcolor{red!10}.752 & .220
& \cellcolor{red!10}.491 &\cellcolor{green!10}.522 & .235&\cellcolor{green!10}.715 & \cellcolor{red!10}.735 & .155 \\
SMV &  \cellcolor{green!10}.495 &\cellcolor{green!10}.513 & .341& \cellcolor{red!10}.735  &\cellcolor{green!10}.759 & .262
& \cellcolor{red!10}.492 &\cellcolor{green!10}.521 & .344&\cellcolor{green!10}.716 &\cellcolor{red!10}.735& .209 \\
NQC & \cellcolor{green!10}.503 & \cellcolor{green!10}.517 & \textbf{.386} &\cellcolor{green!10}.752 & \cellcolor{red!10}.751 & .295
& \cellcolor{green!10}.503&\cellcolor{green!10}.532 & \textbf{.387}&\cellcolor{green!10}.726 &\cellcolor{green!10}.741 &.273\\
UEF & \cellcolor{green!10} .499 & \cellcolor{green!10}.516 & .371 & \cellcolor{green!10}.752 & \cellcolor{red!10}.751  & \textbf{.297}
& \cellcolor{green!10}.500 &\cellcolor{green!10}.531 & .360&\cellcolor{green!10}.721 &\cellcolor{green!10}.740 & .257\\
RSD & \cellcolor{green!10} \textbf{.509} & \cellcolor{green!10}\textbf{.523} & .380& \cellcolor{green!10}.755 & \cellcolor{green!10}\textbf{.770}& .275
& \cellcolor{green!10}\textbf{.517} &\cellcolor{green!10}\textbf{.534} & .386&\cellcolor{green!10}\textbf{.736} &\cellcolor{green!10}\textbf{.745} & \textbf{.278} \\
QPP-PRP & \cellcolor{red!10}.471 & \cellcolor{red!10}.492 & .164 &\cellcolor{red!10}.720 & \cellcolor{red!10}.745& .127
& \cellcolor{green!10}.507 &\cellcolor{green!10}.521 & .255&\cellcolor{green!10}.723 &\cellcolor{red!10}.733 & .116\\
WIG & \cellcolor{red!10} .477  & \cellcolor{red!10}.493 & .223& \cellcolor{red!10}.722 & \cellcolor{red!10}.739& .182
& \cellcolor{red!10}.495 &\cellcolor{red!10}.518 & .061&\cellcolor{green!10}.728 &\cellcolor{green!10}.737& .069 \\
SCNQC & \cellcolor{green!10}.503 & \cellcolor{green!10}.517 & .385 & \cellcolor{green!10}.752 & \cellcolor{red!10}.751 & .295
& \cellcolor{green!10}.503 &\cellcolor{green!10}.532 &  .386&\cellcolor{green!10}.726 &\cellcolor{green!10}.741 &.273\\
QV-NQC & \cellcolor{green!10}.503  & \cellcolor{green!10}.520 & .323& \cellcolor{green!10}\textbf{.757} & \cellcolor{green!10}.764& .244
& \cellcolor{red!10}.494 & \cellcolor{green!10}.521 & .338&\cellcolor{green!10}.715 & \cellcolor{red!10}.717 & .206\\
DM & \cellcolor{green!10}\textbf{.509} & \cellcolor{green!10}.514 & .181 & \cellcolor{green!10}.750  & \cellcolor{green!10}.758 & .111
& \cellcolor{red!10}.496 & \cellcolor{red!10}.514 & .230&\cellcolor{green!10}.715 & \cellcolor{red!10}.725& .184\\
BERT-QPP & \cellcolor{green!10}.504 & \cellcolor{red!10}.498 & .190& \cellcolor{green!10}.741 &  \cellcolor{red!10}.754 & .145
& \cellcolor{green!10} .507& \cellcolor{red!10}.514 & .229&\cellcolor{green!10}\textbf{.736} & \cellcolor{red!10}.733 & .112\\
NQA-QPP & \cellcolor{green!10}.492 & \cellcolor{green!10}.507 & .149& \cellcolor{green!10}.739& \cellcolor{red!10} .752 &.111
& \cellcolor{red!10}.492 & \cellcolor{red!10}.518 & .241&\cellcolor{red!10}.705 & \cellcolor{red!10}.727 & .176\\
\bottomrule
\end{tabular}
\end{adjustbox}
\caption{\small Effectiveness of QPP-based weighted extensions of \rrf~\cite{rrf} and \comb~\cite{fox1994combination} fusion methods on TREC DL'19 and TREC DL'20 topic sets. The best QPP-F values in each column are bold-faced.
Green or red colours indicate results better or worse than the corresponding unweighted fusion-based baseline, respectively.
Additionally, we report the standard correlation based QPP evaluation results, $\standard$, for each QPP model. These correlation values are averaged over the 8 rankers used in our experiments.
}
\label{tab:qpp_fusion_final}
\end{table}

\begin{figure}[t]   
\centering
\begin{subfigure}{0.42\columnwidth}
  \centering
  \includegraphics[width=1\textwidth]{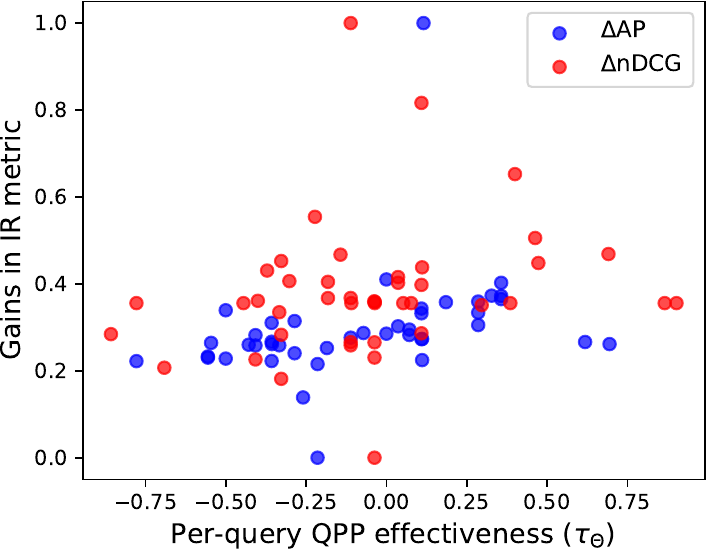}
  \caption{TREC DL'19}
  \label{fig:w/rrf_ap}
\end{subfigure}
\quad
\begin{subfigure}{0.42\columnwidth}
  \centering
  \includegraphics[width=1\textwidth]{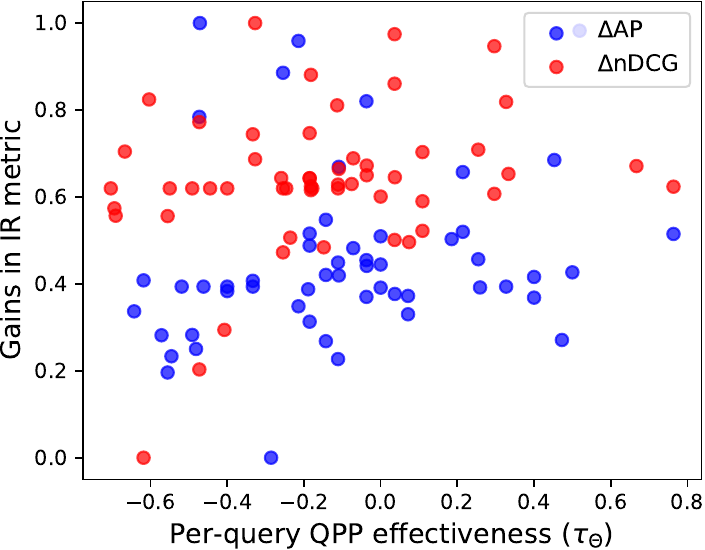}
  \caption{TREC DL'20}
  \label{fig:w/csum_ap}
\end{subfigure}
\caption{\small Correlation between the normalised per-query QPP effectiveness ($\tau_\Theta$) and downstream IR metric improvement ($\Delta_\mathcal{P}$, where $\mathcal{P} \in \{\text{AP@100}, \text{nDCG@10}\}$) for RSD-weighted \comb~(the best performing predictor from Table \ref{tab:qpp_fusion_final}) over unweighted \comb. 
 Pearson’s $\rho$ values are: (a) for DL'19, $\rho = 0.5695$ (AP@100) and $\rho = 0.2789$ (nDCG@10); and (b) for DL'20, $\rho = 0.2876$ (AP@100) and $\rho = 0.2307$ (nDCG@10).
%
}
\label{fig:correlation}
\end{figure}

\section{Results}

\para{Main Observations} In relation to \textbf{RQ1}, Table~\ref{tab:qpp_fusion_final} shows that incorporating QPP predicted scores as weights for fusion (Equation \ref{eq:qppfusion}) consistently improves retrieval effectiveness for most QPP models in both CSUM and RRF setups, e.g., as seen by comparing the unweighted fusion results (\rrf~and \comb) with the corresponding ones obtained with QPP weights along the same column.
QPP estimates work more effectively in conjunction with retrieval scores than reciprocal ranks, as can be seen from the better retrieval results along the \comb~column.

%
%

Among QPP models, the retrieval-score-based unsupervised ones (e.g., NQC, RSD, UEF) consistently outperform more recent content-based supervised approaches (e.g., NQA-QPP, BERT-QPP). In particular, in turns out that RSD-weighted \comb~and \rrf~outperform the weighted fusion derived from other QPP models in terms of both AP@100 and nDCG@10 on both the datasets. In general, it is observed that a majority of the QPP-based weighted fusion approach fail to improve nDCG@10 over its unweighted counterpart except the score-based ones which mostly perform well even for nDCG@10.



We additionally report the standard QPP evaluation measure of how correlated (Kendall's $\tau$) the estimates are with the respective target metric (AP or nDCG).
In relation to \textbf{RQ2}, Table~\ref{tab:qpp_fusion_final} shows the best downstream-agnostic performance of the predictors ($\standard$) often does not align with the IR fusion (downstream-aware) results, e.g., NQC yields the best correlation with AP on DL'19, whereas these estimates do not lead to the best downstream use. Specifically, the correlation between the relative ordering of the QPP models between downstream retrieval effectiveness of weighted \comb~fusion and  $\standard$ evaluation (see Section \ref{ss:setup}) for AP is $\tau=0.4805$ for DL'19 and $\tau = 0.6798$ for DL'20.

\para{Per-query Analysis}

In relation to \textbf{RQ3}, Figure \ref{fig:correlation} shows the per-query correlations of predictions across the different ranking models for a particular query against the downstream IR effectiveness gains obtained on that query. To avoid clutter, for this analysis we present results obtained only with the best performing QPP model, namely RSD, as seen from Table \ref{tab:qpp_fusion_final}.


We observe a \textbf{moderately positive correlation} (values reported in the caption of Figure \ref{fig:correlation}) between the per-query effectiveness of ranker prediction and the downstream fusion gains indicating that better fusion results may be achieved even with some error in predicting the relative ranker preference. 


\section{Concluding Remarks}
In this work, we present a downstream-aware query performance prediction (QPP) evaluation that first involves predicting the relative preference between an available set of ranking models on a per-query basis, and then eventually uses these estimates as priors in a fusion-based IR pipeline.
The effectiveness of a QPP model is then measured by how effectively do the per-query ranking model preference estimates when used as fusion priors contribute to an overall increase of retrieval effectiveness.
Our experiments not only demonstrate that
incorporating QPP estimates as priors in weighted fusion mostly lead improving IR effectiveness in terms of AP@100 and nDCG@10 on both TREC DL'19 and '20, but also reveal a new way of evaluating QPP models. In particular, we show that this evaluation differs considerably from the standard practice of QPP evaluation in that the downstream evaluation yields a different relative preferential ordering of the QPP systems.

In future, we plan to develop novel QPP approaches for addressing this fine-grained task of predicting the best ranker on a per-query basis.  Moreover, we also plan to apply the QPP estimates for developing an adaptive pipeline for RAG systems.





\subsubsection*{Disclosure of Interests}
We declare that we don't have any competing interests.

\bibliographystyle{splncs04}
\bibliography{references.bib}
\end{document}